# Nonlinear mechanics of rigidifying curves

Salem Al Mosleh and Christian Santangelo*

*Physics Department, University of Massachusetts Amherst, Amherst, MA 01003, USA*

Thin shells are characterized by a high cost of stretching compared to bending. As a result isometries of the midsurface of a shell play a crucial role in their mechanics. In turn, curves on the midsurface with zero normal curvature play a critical role in determining the number and behavior of isometries. In this paper, we show how the presence of these curves results in a decrease in the number of linear isometries. Paradoxically, shells are also known to continuously fold more easily across these rigidifying curves than other curves on the surface. We show how including nonlinearities in the strain can explain these phenomena and demonstrate folding isometries with explicit solutions to the nonlinear isometry equations. In addition to explicit solutions, exact geometric arguments are given to validate and guide our analysis in a coordinate-free way.

## I. INTRODUCTION

As a thin, elastic structure is deformed, it tends to flex without appreciably stretching. This is evident in a sheet of paper for example, which is soft to bending deformations but highly resistant to stretching [1]. Even under significant deformation, thin elastic structures tend to concentrate their stretching distortion into small regions of high strain surrounded by bent but relatively unstretched regions [2–6]. Because of this, isometries – deformations that deform a surface without stretching – play a privileged role in the mechanics of thin shells.

Roughly speaking, the more isometric deformations there are, the more ways you can deform a shell without stretching it. For example, the cross-sectional geometry of a thin, cylindrical shell can be deformed easily whereas a complete spherical shell cannot without introducing in-plane, and elastically costly, stretching. Indeed, a closed surface, such as a sphere, generically has no infinitesimal, smooth isometries [8]. Any deformations of a spherical shell must, therefore, balance stretching and bending.

Interestingly, Tenenblat [9] and, later, Audoly [10, 11], pointed out that in the vicinity of asymptotic curves– curves with zero normal curvature – the infinitesimal isometry equations are singular. As a result, there will be fewer smooth isometric degrees of freedom near those curves. Does that mean that these surfaces are more rigid in the zero thickness limit? If so, that would seem to be at odds with experiments described in Ref. [12], in which it was shown that a shell can be folded continuously across an asymptotic curve without any stretching at all, while folding across a non-asymptotic curve would require traversing a stretching energy barrier.

In order to reconcile these two observations we need to find a (not necessarily smooth) family of nearly isometric deformations that connects the undeformed and folded states. And show that this deformation is energetically favored in the experimental conditions of [12]. This last point is discussed further in the conclusion in Sec. IV.

In this paper, we seek to resolve this potential difficulty by accounting for the linearities in the elastic strain. Like the bending energy, these nonlinearities can also regularize the divergences in the linear isometries and naturally lead to the folded solution. We estimate the thickness range for which the correction due to the nonlinearities will be dominant over the bending energy considered in [10].

The paper is organized as follows. In section II, we give an overview of linear isometries and discuss their existence and properties. We derive the infinitesimal isometry equations and give explicit solutions to the particular case of a parabolic torus (Eq. 18), the behavior of the smooth and diverging solutions are explored near the rigidifying curves. In section III, we show how the addition of the nonlinear terms in the isometry equation can regularize the divergences. We derive an approximate solution using the tools of boundary layer theory. The nonlinear solutions are then used to explain how folding across a rigidifying curve happens continuously and isometrically. We conclude in section IV.

## II. LINEAR ISOMETRIES FOR AXISYMMETRIC SURFACES

### A. Isometric deformations and mechanics of shells

We start this section by giving an overview of the relationship between isometric deformations and the mechanics of thin shells.

Starting with an undeformed shell, there are two related considerations for understanding the role of isometries in the mechanics of shells. First, the allowed isometric deformations may be smooth or non-smooth. For example a sphere admits $C^1$ isometries with infinite bending cost. Of course in a real shell the sharp feature will be smoothed out leading to finite energy cost related to the shell thickness in a nontrivial way [4]. A cylinder, on the other hand, admits many smooth isometries [7]. In principle, there can also be nonsmooth isometries with better continuity than $C^1$ [13, 14]. These isometries are far less costly even for small thickness.

---

*e-mail: csantang@physics.umass.edu

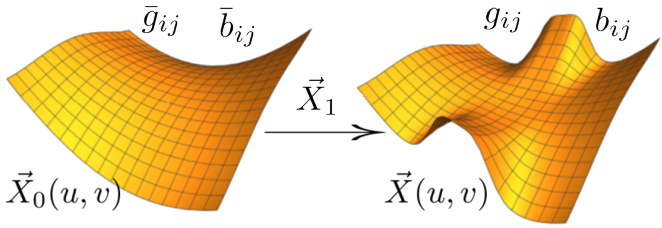

FIG. 1: Displacement vector between two points on the mid-surface of a shell.

Second, one can ask whether an isometry is connected to the undeformed state by a continuous one parameter family (or families) of isometries $\mathbf{X}(u^1, u^2, \epsilon)$. It was shown in the experiments in Ref. [12] that a shell can be continuously folded across an asymptotic curve without snapping, implying the existence of a family of isometric deformations connecting the undeformed and folded states.

For a smooth family $\mathbf{X}(u^1, u^2, \epsilon)$, we may define the infinitesmal isometry as $\mathbf{X}_1(u^1, u^2) \equiv \partial_\epsilon \mathbf{X}(\epsilon = 0)$. Equivalently, we may write

$$\mathbf{X}_0(u^1, u^2, \epsilon) \approx \mathbf{X}_0(u^1, u^2) + \epsilon\, \mathbf{X}_1(u^1, u^2). \tag{1}$$

If a smooth family of isometries $\mathbf{X}_0(u^1, u^2, \epsilon)$ fails ot exist, which implies the absence of infinitesimal isometries, the surface is said to be "rigid" in the sense defined in Ref. [8], in which it was proved that almost all simply connected closed surfaces are rigid.

### B. General Linear Isometries and Self-Stresses

To be self-contained and to establish our notation, we start with a review of linear (or infinitesimal) isometries. We parametrize the shape of the shell in terms of the coordinates, $\mathbf{u} = (u^1, u^2)$, of its mid-surface (Fig. 1). We start with a reference surface, who's shape is given by the three-dimensional position of each point through a vector function $\mathbf{X}_0(u^1, u^2)$. Therefore, the preferred distance between two points described by coordinates $u^i + du^i$ and $u^i$, for infinitesimal $du^i$, is given by the first fundamental form,

$$d\mathbf{X}_0^2 = \partial_i \mathbf{X}_0 \cdot \partial_j \mathbf{X}_0\, du^i du^j \equiv \bar{g}_{ij}\, du^i du^j \tag{2}$$

where, in accordance with the Einstein summation convention, repeated indices are summed unless explicitly stated. The last equality defines the components of the (induced) reference metric tensor $\bar{g}_{ij}$, which encodes the equilibrium distances on the surface and must be symmetric and positive-definite. Similarly we define a deformed metric $g_{ij}$ for the deformed surface $\mathbf{X}(u^1, u^2)$. Deformations for which $g_{ij} = \bar{g}_{ij}$, called isometries, satisfy

$$\partial_i \mathbf{X} \cdot \partial_i \mathbf{X} = \bar{g}_{ij}. \tag{3}$$

Consider a curve on the surface, with space curvature $\kappa$. The normal curvature is the projection of the curvature vector along the normal to the surface, and the geodesic curvature is the projection along the tangent plane. This naturally leads to the relation

$$\kappa^2 = \kappa_N^2 + \kappa_g^2. \tag{4}$$

Curves with zero normal curvature are called asymptotic curves. Interestingly, the geodesic curvature does not change under isometric deformations, we exploit this fact in appendix A. For an arc length parametrization of the curve $u^i(s)$, the normal curvature is given by

$$\kappa_N = (\hat{\mathbf{N}} \cdot \partial_i \partial_j \mathbf{X}) \frac{du^i}{ds} \frac{du^i}{ds}, \tag{5}$$

where $\hat{\mathbf{N}}$ is the normal to the surface. We define the expression in the parentheses as the curvature tensor $b_{ij} \equiv \hat{\mathbf{N}} \cdot \partial_i \partial_j \mathbf{X}$.

For shells of very small thickness compared to curvature, we generically expect the deformations to be dominated by isometries [1]. This propensity is characterized by the Föppl-von Kàrmàn number, FvK = $BR^2/Y$, which measures the ratio of the bending stiffness $B$, characteristic length $R$ and Young's modulus $Y$ [15]. Typically, FvK $\propto R^2/t^2$ for shells of thickness $t$, showing that in-plane elasticity dominates over any bending energies [16, 17]. For large FvK, we study the deformations of a shell using the in-plane elastic energy [18]

$$E_s = \frac{1}{2} \int dA\, T^{ij} \left( \partial_i \mathbf{X} \cdot \partial_j \mathbf{X} - \bar{g}_{ij} \right), \tag{6}$$

where the stress $T^{ij}$, a symmetric tensor, is treated as a Lagrange multiplier to force the deformation to lie along an isometry. To this we add a bending energy

$$E_b = \frac{B}{2} \int dA\, \left( b_{ij} - \bar{b}_{ij} \right) \left( b^{ij} - \bar{b}^{ij} \right), \tag{7}$$

where $\bar{b}_{ij}$ measures the intrinsic curvature of the shell.

Next we derive equations governing the isometries of a shell's midsurface, $\mathbf{X}_0$. Consider a small deformation $\mathbf{X} = \mathbf{X}_0 + \mathbf{X}_1$ and a corresponding deformation of $T^{ij}$ to $T^{ij} + T_1^{ij}$. Substituting this into the in-plane elastic energy and expanding to lowest order, we obtain

$$\begin{aligned} \delta E_s = & -\int dA\, \left( D_i T^{ij} \partial_j \mathbf{X}_0 \cdot \mathbf{X}_1 + T^{ij} \bar{b}_{ij} \hat{\mathbf{N}}_0 \cdot \mathbf{X}_1 \right) \\ & + \int dA\, T_1^{ij} \left( \partial_i \mathbf{X}_0 \cdot \partial_j \mathbf{X}_1 \right) \\ & + \oint d\ell\, T^{ij} \hat{\mathbf{n}}_i \partial_j \mathbf{X}_0, \end{aligned} \tag{8}$$

where $D_i$ is the covariant derivative with respect to $\mathbf{X}_0$, $\hat{\mathbf{n}}$ is a vector tangent to the midsurface but normal to the boundary, and $d\ell$ is the integral over the boundary with respect to arc length.

Decomposing $\mathbf{X}_1$ into components tangent and normal to the surface,

$$\mathbf{X}_1 = A_n(u^1, u^2)\, \hat{\mathbf{N}} + A_i(u^1, u^2)\, \hat{\mathbf{e}}^i, \qquad (9)$$

where $\hat{\mathbf{e}}^i$ are vectors tangent to the surface satisfying $\partial_i \mathbf{X}_0 \cdot \hat{\mathbf{e}}^j = \delta_i^j$ and $\delta_i^j$ is the Kronecker delta, we find that, to linear order, an isometry satisfies

$$-2\, \bar{b}_{ij} A_n(u^1, u^2) + D_i A_j(u^1, u^2) + D_j A_i(u^1, u^2) = 0 \qquad (10)$$

while the stress satisfies

$$D_i T^{ij} = 0, \quad T^{ij} \bar{b}_{ij} = 0, \qquad (11)$$

subject to the boundary condition $T^{ij} \hat{\mathbf{n}}_i = 0$. We call any $\mathbf{X}_1$ that satisfies Eq. (10) a *first-order*, or infinitesimal, isometry and any nonzero solution of Eqs. (11) a *self-stress*. The relationship between self-stresses and isometries can be understood by index theory, but this is outside the scope of the current paper [19].

When can we find a solution to the three equations in Eq. (10)? Naively, $A_n$ appears algebraically in Eq. (10) and can be eliminated, leaving two equations in two unknowns. Since Eq. (10) is first-order, it appears that specifying the two in-plane deformations of a surface along a single curve is sufficient to determine the isometric deformation of the entire surface uniquely. However, the Gaussian curvature of the surface itself determines whether Eqs. (10) are elliptic or hyperbolic [1, 10]. Thus, any curve along which the Gaussian curvature changes sign, or alternatively one of the principle curvatures changes sign as it does in the torus (Fig. 2), changes the character of the isometry equations. Specifically, asymptotic curves (where $\kappa_N \equiv 0$) are the characteristics of Eq. (10). Because information propagates along these curves, unlike other curves, arbitrary boundary conditions cannot be specified on them. In other words, they have fewer infinitesimal isometric degrees of freedom. A consequence of the change from elliptic to hyperbolic in Eqs. (10) is that some linear isometries appear to diverge as they approach the curves of $\kappa_N, K = 0$ [10, 11].

Even when the Gaussian curvature does not vanish, a problem with Eqs. (10) can develop. For example, the horizon at the pseudosphere (Fig. 3) has one vanishing and one diverging principle curvature and cannot, consequently, be extended beyond its boundary despite K being constant.

### C. Axisymmetric Surfaces and rigidifying curves

To demonstrate these features in simplest context, we specialize the linear isometry equations to axisymmetric surfaces. In that case, the embedding, $\vec{X}_0$, can be expressed as

$$\vec{X}_0(s, \theta) = s\, \hat{s}(\theta) + h(s)\, \hat{z}, \qquad (12)$$

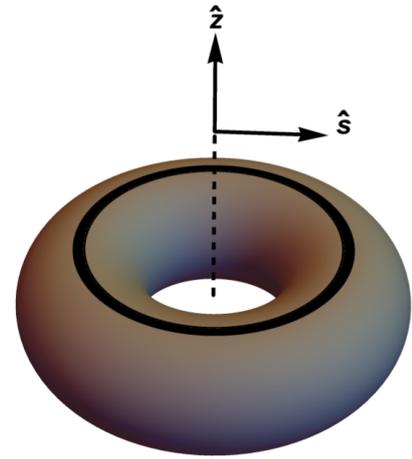

FIG. 2: Torus with a curve of $\kappa_N = 0$ shown in black.

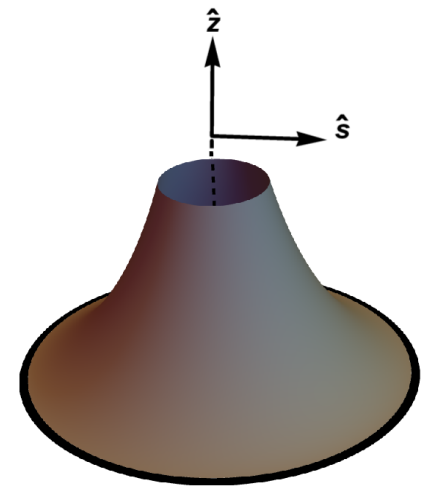

FIG. 3: Pseudosphere with the curve of $\kappa_N = 0$ shown in black.

where $s$ is the radial distance from the $z-$axis, $\hat{s}$ is the unit vector pointing in the radial direction, and $h(s)$ is the vertical height of the surface. The tangent vectors are $\partial_\theta \vec{X}_0 = s\, \hat{\theta}$ and $\partial_s \vec{X}_0 = \hat{s} + h'(s)\, \hat{z}$, where the prime indicates a derivative with respect to $s$. The first and second fundamental forms are

$$\bar{g}_{ij} = \begin{pmatrix} 1 + h'(s)^2 & 0 \\ 0 & s^2 \end{pmatrix} \qquad (13)$$

and

$$\bar{b}_{ij} = \frac{1}{\sqrt{1 + h'(s)^2}} \begin{pmatrix} h''(s) & 0 \\ 0 & s\, h'(s) \end{pmatrix}. \qquad (14)$$

The normal curvature along the curves of constant $s$ vanishes when $h'(s) = 0$; these are precisely the rigidifying curves.

An arbitrary displacement of the surface can be written

as

$$\vec{X}_1(s,\theta) = A_s(s,\theta)\hat{s}(\theta) + A_\theta(s,\theta)\hat{\theta}(\theta) + A_z(s,\theta)\hat{z} \quad (15)$$

in terms of the basis $(\hat{s}, \hat{\theta}, \hat{z})$. We can exploit the axisymmetry of the isometry equations by expressing them in terms of the Fourier transforms of the functions $A_i(s,\theta)$,

$$A_i(s,\theta) = \sum_m \tilde{A}_i(s,m)\, e^{im\theta} \quad (16)$$

where $m$ is an integer. After some algebra, these three equations can be combined into a single equation for each mode $m$ for $\tilde{A}$,

$$\tilde{A}_z''(s) + \frac{h''(s)}{h'(s)} \tilde{A}_z'(s) - \frac{m^2\, h''(s)}{s\, h'(s)} \tilde{A}_z(s) = 0. \quad (17)$$

This second-order differential equation has two analytic solutions as long as $h'(s) \neq 0$. If there is a singularity, $h'(s^*) = 0$ for some $s = s^*$, it will be regular if $h'(s) \sim (s - s^*)$ as $s$ approaches $s^*$.

We illustrate the behavior of the isometries near the curve $h'(s^*) = 0$, by considering the "parabolic torus" near the rigidifying curve described by

$$h(s) = \frac{1}{2\,a}\,(s - R)^2. \quad (18)$$

The surface of Eq. (18) can be thought of as an approximation of more general axisymmetric surfaces near a rigidifying curve at $s = s^*$. Since $h'(R) = 0$ with $h''(R) \neq 0$, Eq. (17) has a regular singularity at $s = R$ and, indeed, the surface described by $h(s)$ has a circle of zero normal curvature along $s = R$. In dimensionless variables, the linear isometry equation becomes

$$\partial_u^2 Y_z(u) + \frac{1}{u} \partial_u Y_z(u) - \frac{m^2}{u(u+1)} Y_z(u) = 0, \quad (19)$$

where $u \equiv (s - R)/R$ and $\tilde{A}_z(u) \equiv R\, Y_z(u)$. The solutions have the form

$$Y_{zm} = A_m\, {}_2F_1(-m, m; 1; -u) + B_m\, \log|u|\, {}_2F_1(-m, m; 1; -u) + B_m\, {}_2G_1(-m, m; 1; -u), \quad (20)$$

and

$${}_2G_1(\alpha, \beta; 1; -u) \equiv \sum_{r=1}^{\infty} \left[ \frac{\Gamma(\alpha + r)\, \Gamma(\beta + r)}{\Gamma(\alpha)\Gamma(\beta)(r!)^2} \sum_{k=0}^{r-1} \left( \frac{1}{\alpha + k} + \frac{1}{\beta + k} + \frac{2}{1+k} \right) (-u)^r \right], \quad (21)$$

where $A_m$ and $B_m$, are constants, ${}_2F_1(a, b; c; z)$ is a hypergeometric function, and we have defined the analytic function ${}_2G_1(-m, m; 1; -u)$. The functions $A_{zm}(u)$, $A_{sm}(u)$ and $A_{\theta m}(u)$, for $m > 1$, can be found from $Y_{zm}(u)$ and the isometry equations. Figures (4 -7) illustrate these solutions for the parabolic torus of Eq. (18).

Thus, Eq. (20) has one solution that diverges logarithmically as $u \to 0$ for each mode $m$. The asymptotic behavior of the solutions for $u \ll 1$ is

$$Y_{zm}(u) = A_m + B_m\, \log|u| + O(u \log u). \quad (22)$$

Taking the inverse Fourier transform we can rewrite this limit as

$$Y_z(u, \theta) = A(\theta) + B(\theta)\, \log|u| + O(u \log u) \quad (23)$$

where $A(\theta)$ and $B(\theta)$ are arbitrary functions of $\theta$. Any axisymmetric surfaces satisfying $h(s) \propto (s - s^*) + O\left((s - s^*)^2\right)$ will have the same leading behavior given in Eq. (23). In the next section, this form will be convenient for asymptotic matching to the nonlinear solution in the vicinity of the rigidifying curve. The solutions to $A_\theta$ and $A_s$ corresponding to the diverging solution are also non-analytic at $u = 0$, but they both vanish as $O(u \log u)$ as $u \to 0$. Naive considerations would suggest that we require $B_m = 0$ to avoid the divergences that occur in the isometry. However, the approximation of linear elasticity also breaks down near the rigidifying curve. As we will see in the next section, when we include nonlinear terms in our analysis, the divergence of the isometry is regularized.

### III. NONLINEAR MECHANICS OF RIGIDIFYING CURVES

Though we may be tempted to exclude the diverging solutions, only the vicinity of the rigidifying curves becomes rapidly varying and large. Thus, two of our assumptions become invalid near the rigidifying curves: the bending energy may not be neglected and geometrical nonlinearities in the strain are no longer negligible. For sufficiently thin surfaces (see Eq. 28), the bending energy can always be made smaller than the nonlinear strain terms, therefore we will consider the effect of the unavoidable nonlinearities.

The full nonlinear isometry equations are, unfortu-

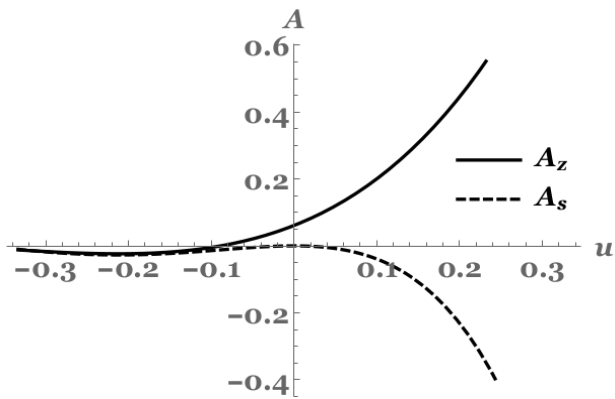

FIG. 4: Smooth isometric deformations of the parabolic torus, normalized so that A(s = R + a) = 1 and m = 4. $A_z$ and $A_s$ represent displacements of the initial surface, rather than absolute positions.

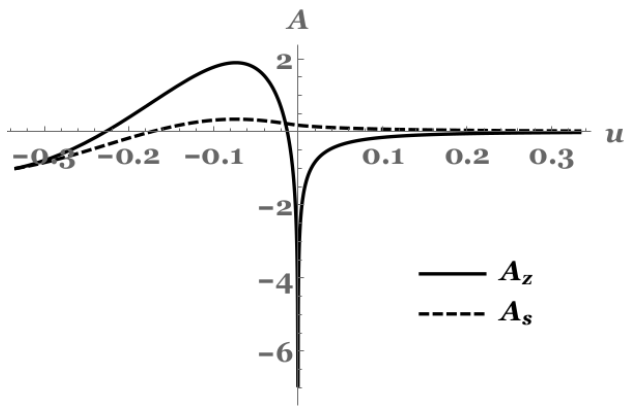

FIG. 5: Diverging isometric deformations of the parabolic torus, normalized so that A(s = R - a) = 1 and m = 4. Notice that the displacement in the z-direction behaves as $A_z \sim \log u$ as $u \to 0$

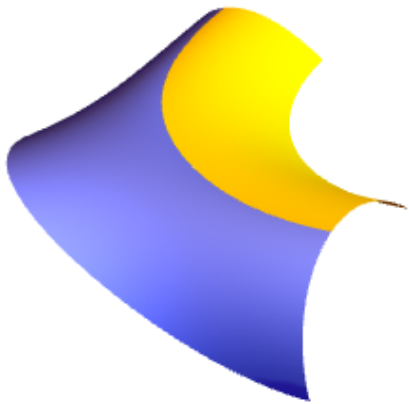

FIG. 6: Surface deformed by smooth linear isometry.

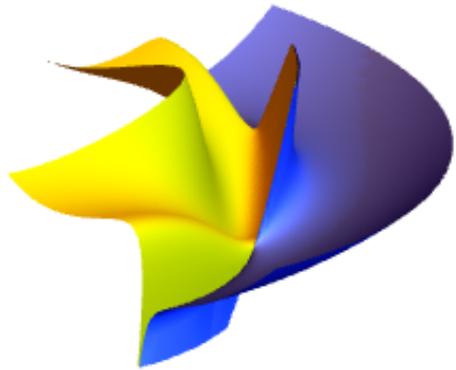

FIG. 7: Surface deformed by diverging linear isometry.

nately, complicated. They read

$$0 = 2\partial_s A_s + 2h'(s)\partial_s A_z + (\partial_s A_s)^2 \quad (24)$$
$$+(\partial_s A_\theta)^2 + (\partial_s A_z)^2,$$

$$0 = 2s(\partial_\theta A_\theta + A_s) + (\partial_\theta A_s - A_\theta)^2 \quad (25)$$
$$+(\partial_\theta A_\theta + A_s)^2 + (\partial_\theta A_z)^2,$$

and

$$0 = s\partial_s A_\theta + \partial_\theta A_s - A_\theta + h'(s)\partial_\theta A_z$$
$$+ (\partial_\theta A_s - A_\theta)\partial_s A_s + \quad (26)$$
$$(\partial_\theta A_\theta + A_s)\partial_s A_\theta + (\partial_\theta A_z)(\partial_s A_z).$$

As in section II C, we assume $h(u) \approx (s-R)^2/(2a)$ and $u = (s-R)/R$ as $s \sim R$.

If we substitute in the linear solution, we note that $A_\theta \sim u \log u$ and $A_s \sim u \log u$, suggesting that the terms nonlinear in $A_s$ and $A_\theta$ can be ignored. This approximation can be justified post-hoc. Within this approximation, we set $A_z(s, \theta) = R^2 Y_z(u, \theta)/(2a)$ and use the linearized forms of Eqs. (24) and (25) to eliminate $A_s$ and $A_\theta$. Thus, we obtain a single equation for $Y_z$,

$$0 = 2(1+u)\partial_u Y_z + 2u(1+u)\partial_u^2 Y_z + 2\partial_\theta^2 Y_z$$
$$-\frac{[\partial_\theta Y_z - (1+u)\partial_\theta \partial_u Y_z]^2}{(u+1)^2} + \quad (27)$$
$$[\partial_\theta^2 Y_z + (1+u)(2u + \partial_u Y_z)]\partial_u^2 Y_z.$$

Eq. (27) can be solved numerically (the results are shown in Appendix B). Here, we will pursue an analytic approach to obtaining approximate solutions. There are three regimes. The linear solution is valid when $u \gg |Y_z(u \sim \frac{a}{R})|^{1/2}$. Within the layer $u \lesssim |Y_z(u \sim \frac{a}{R})|^{1/2}$, the nonlinearities become important and the linear solution is no longer valid. For the nonlinearities to become important before the bending energy modifies the solution, the width of this layer ($\lambda_N$) must be bigger than the width of the layer that would result from bending

energy regularization $\lambda_B \sim (t\,R\,a)^{1/3}$ [11]. Therefore, our analysis is valid when

$$\lambda_B \ll \lambda_N \implies t \ll \frac{R^2 \, |Y_z(u \sim \frac{a}{R})|^{3/2}}{a}. \tag{28}$$

When this condition is met we can obtain a finite solution to Eq. (27) in powers of $u$ near the rigidifying curve (Appendix B). This shows that the nonlinear terms are sufficient to regularize the divergences of the nonlinear theory. This is the "inner solution" in the language of boundary layer theory. In the intermediate regime, $|Y_z(u \sim \frac{a}{R})|^{1/2} \ll u \ll 1$, we may obtain a better approximation by considering how large the various terms in the nonlinear equation become as we approach the rigidifying curve when substituting the linear solution into Eq. (27). The most divergent term is proportional to $\partial_u Y_z \, \partial_u^2 Y_z$, which behaves as $\sim 1/u^3$ for small $u$, whereas all the other nonlinear terms diverge as $\sim u^{-2}$ or slower. On the other hand, the first two linear terms in the equation are $O(u^{-1})$ and the linear term $\partial_\theta^2 Y_z$ only grows as $\log u$.

Taken together, this suggests that Eq. (27) has a regime, $|Y_z(u \sim \frac{a}{R})|^{1/2} \ll u \ll 1$, where all the nonlinear terms except the last term can be treated as a perturbation. This argument does not work when the coefficient of the $\log u$ term vanishes at some value of $\theta$. We will treat these regions separately and unless otherwise stated we will assume that the coefficient of the $\log u$ solution is greater than zero.

The resulting reduced equation can be written in the form

$$\partial_u \left[ u \partial_u Y_z + \frac{1}{4}(\partial_u Y_z)^2 \right] = 0, \tag{29}$$

and solved by

$$\partial_u Y_z(u, \theta) = -2\,u \pm \sqrt{4\,u^2 + \gamma_\pm(\theta)}, \tag{30}$$

where $\gamma(\theta)$ is a constant of integration. We can now solve for the z-component of the displacement by integrating (30) to find

$$Y_{z\pm}(u) = \delta_\pm(\theta) - u^2 \pm \frac{u}{2}\sqrt{4\,u^2 + \gamma_\pm(\theta)} + \frac{\gamma_\pm(\theta)}{8} \log \left( \frac{\pm 2\,u + \sqrt{4\,u^2 + \gamma_\pm(\theta)}}{\sqrt{\gamma_\pm(\theta)}} \right)^2, \tag{31}$$

where $\delta_\pm(\theta)$ is another integration constant. Note that there are two branches of solution, shown in figures (8), with opposite signs of the normal curvature $\kappa_N \sim \pm\sqrt{\gamma_\pm}$. This is what we would expect (see [20]) from the relation $\kappa_N = \pm \sqrt{\kappa^2 - \kappa_g^2} \sim \pm \sqrt{\delta \kappa}$ (see Eq. (4) and appendix A for more details).

Note that $Y_{z\pm}$ remains finite as $u \to 0$, quite unlike it does in the case of infinitesimal isometries. In the limit $u/|\gamma_\pm|^{1/2} \ll 1$, this solution behaves as the regular series

$$Y_{z\pm} = \delta_\pm(\theta) \pm \sqrt{\gamma_\pm(\theta)} - u^2 \pm \frac{2\,u^3}{3\sqrt{\gamma_\pm}}, \tag{32}$$

This series can be matched to a series expansion solution of Eq. (27) near the rigidifying curve. Notice that the solution does not make sense unless $\gamma_\pm > 0$. In light of the relation $\kappa_N \sim \pm \sqrt{\delta \kappa} \sim \pm\sqrt{\gamma_\pm}$, this requirement is equivalent to the requirement that $\delta \kappa > 0$, or that $\kappa > \kappa_g$ within our approximation. More generally it can be shown that the full (no approximations) isometry equations are not well behaved when $\kappa_N$ changes sign (see appendix A).

The limit $u/|\gamma_\pm|^{1/2} \gg 1$ (for $u > 0$), on the other

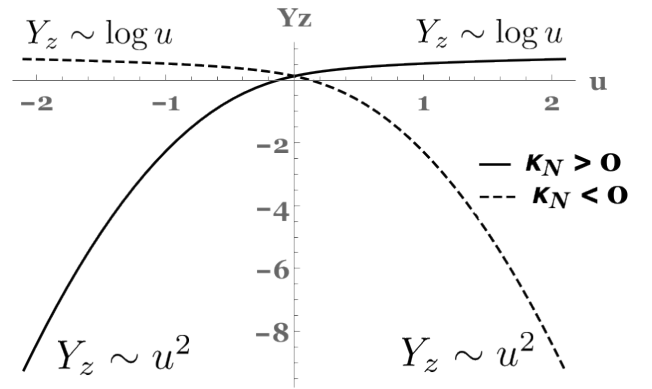

FIG. 8: The two inner solutions obtained from equation (III) with $\delta_\pm = 0.2 \cos(2\,\theta)$ and $\gamma_\pm = 0.6 + 0.3 \cos(2\,\theta)$, evaluated at $\theta = \pi/7$. The two solutions have opposite signs of normal curvature, which is required for existence of folding isometries [12, 20].

hand, yields

$$Y_{z+} = F_+(\theta) + \frac{\gamma_+(\theta)}{4} \log u \qquad (33)$$

$$Y_{z-} = -2u^2 + F_-(\theta) - \frac{1}{4}\gamma_-(\theta) \log u \qquad (34)$$

$$F_\pm = \frac{1}{8}\left(\pm\gamma_\pm(\theta) + \gamma_\pm(\theta) \log\left(\frac{16}{\gamma_\pm(\theta)}\right) + 8\,\delta_\pm(\theta)\right). \qquad (35)$$

The form of this solution can be matched to the infinitesimal isometry far from the rigidying curve. Notice that it is only possible to match one of the solutions ($Y_{z+}$) to the region $u \gg \epsilon^{1/2} > 0$. The $-2u^2$ behavior of $Y_{z-}$ corresponds to the deformation $h(s) \to -h(s)$, which is a reflection of the undeformed surface about the z-axis.

Similarly, if $u < 0$, the limit $|u|/|\gamma_\pm|^{1/2} \gg 1$ yields

$$Y_{z+} = -2u^2 + G_+(\theta) - \frac{1}{4}\gamma_-(\theta) \log u \qquad (36)$$

$$Y_{z-} = G_-(\theta) + \frac{\gamma_+(\theta)}{4} \log u \qquad (37)$$

$$G_\pm = \frac{1}{8}\left(\mp\gamma_\pm(\theta) + \gamma_\pm(\theta) \log\left(\frac{16}{\gamma_\pm(\theta)}\right) + 8\,\delta_\pm(\theta)\right). \qquad (38)$$

We again find that only one of the solutions can match the linear behavior on the $u < 0$ side. Interestingly, in this case it is $Y_{z-}$ that matches the linear solution.

Thus, each of the smooth solutions match the linear isometries only on one side of the rigidifying curve. On the other side, the smooth isometry approximates the reflection of the original parabolic torus about the z-axis. Take for example $A_{z+}$, the z-component of the isometry corresponding to $Y_{z+}$. In the limit $\gamma_+ \to 0$ and $\delta_+ \to 0$ it is approximately equal to

$$A_{z+} = \begin{cases} \frac{(s-R)^2}{2\,a} & s > R \\ \frac{-(s-R)^2}{2\,a} & s < R, \end{cases} \qquad (39)$$

which is the original parabolic torus with the region $s < R$ reflected along the z-axis. While being isometric to the original surface, it is not connected to it through a small displacement. Thus there will be an energy barrier preventing the solutions $Y_{z+}$ and $Y_{z-}$ from being realized starting from the undeformed torus.

To construct a solution that is connected to the parabolic torus through a small displacement, we need to glue $A_{z+}$ in the region $s > R$ with $A_{z-}$ on $s < R$. This results in an isometry that is not smooth on the curve s = R, on one side of the curve, $A_{z+}$ has a positive normal curvature, while the opposite side has a negative normal curvature. Which is what you would expect when two surfaces are joined isometrically along a fold (see Refs [20], [21] and [12]).

Whether this actually happens in practice will depend on the energetics of stretching and bending. The folded solution can be made energetically favorable if the surface is creased (made thinner) at the curve s = R as was done experimentally in [12]. In the remainder of this section, we will construct explicitly the global isometric solution by gluing together the partial solutions in the various regimes.

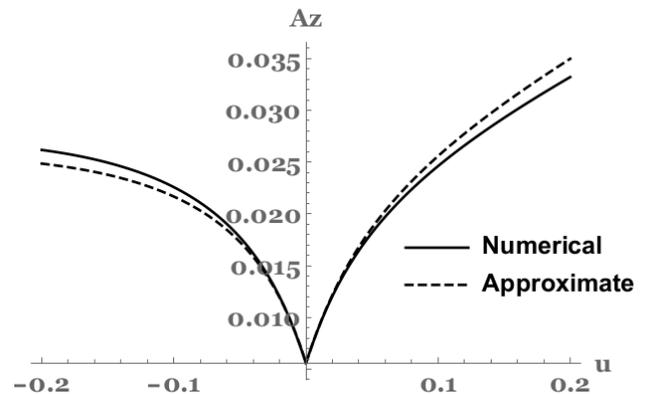

FIG. 9: Comparing the numerical solutions of Eq. (27) to the matched approximation. Here a = 1, R = 3, $\delta_+ = 0.002 \cos(2\,\theta)$ and $\gamma_\pm = 0.006 + 0.003 \cos(2\,\theta)$

We have already seen that in the regime $u \ll 1$, the linear solution takes the form of Eq. (23). There is an overlap between the regions of validity of both these approximations, namely $\delta_\pm^{1/2} \lesssim u \lesssim 1$, so we may match them to obtain [23]

$$A(\theta) = \frac{R}{2\,a} F_+(\theta), \quad B(\theta) = \frac{R}{8\,a} \gamma_+(\theta), \quad u > 0, \quad (40)$$

$$A(\theta) = \frac{R}{2\,a} G_-(\theta), \quad B(\theta) = \frac{R}{8\,a} \gamma_-(\theta), \quad u < 0. \quad (41)$$

Therefore the two assumptions that we made, $B(\theta) \neq 0$ and $\gamma_\pm > 0$, are consistent. Fig. 9 shows the agreement between the matched inner and linear approximations [26] and the numerical solution of Eq. (27).

Consider the setup shown in Fig. 10. We attach a frame of a given shape to the rigidifying curve. Since the curve has only two isometric degrees of freedom, we need only specify $A_z(\theta)$ and $A_s(\theta)$ on the curve and $A_\theta(\theta)$ will be determined. Specifically

$$A_z(s = R, \theta) = \frac{R^2\,\delta_\pm(\theta)}{2\,a} \qquad (42)$$

$$A_s(s = R, \theta) = \frac{R^2\,\tau_\pm(\theta)}{2\,a}, \qquad (43)$$

where $\tau_\pm$ is determined by $\gamma_\pm$ using the series solution of the isometry equation (24), and where we neglected the nonlinearities in $A_\theta$ and $A_s$. To leading order in $|\delta_\pm|$ and up to rigid xy-translations, we have

$$\tau_\pm(\theta) = \frac{a}{R} \operatorname{Re}\left[i\,e^{i\,t} \int_0^t e^{-i\,\sigma}\,\gamma_\pm(\sigma)\,d\sigma\right], \qquad (44)$$

This choice fixes $A_s(u = 0, \theta = 0) = 0$.

Interestingly there is no way to distinguish whether we are in the $\kappa_N > 0$ branch or the $\kappa_N < 0$ branch just by knowing the shape of the deformation at $s = R$. Either isometry can be attached to a given boundary condition and can be continuously reached from the undeformed torus (but not smoothly, because of the $\gamma_\pm^{1/2}$ in the solutions). This is consistent with the results in Refs [12] and [21].
Finally, notice that the constraint $\gamma_\pm > 0$ can be expressed as a constraint on the s-displacement of the rigidifying curve or, equivalently, on the shape of the boundary curve. Consider the curves perpendicular to the rigidifying curve. On an undeformed (parabolic) torus they are (parabolas) circles satisfying $\theta = $ constant, and with tangents at $u = 0$ pointing toward the center of the torus in the xy-plane. After isometric deformation, the tangents are still perpendicular to the rigidifying curve (Fig. 10), but since $\partial_u A_z > 0$ when $u > 0$ and $\partial_u A_z < 0$ when $u < 0$, these curves will be pointing strictly above the xy-plane. This is another way to express the requirement $\gamma > 0$.

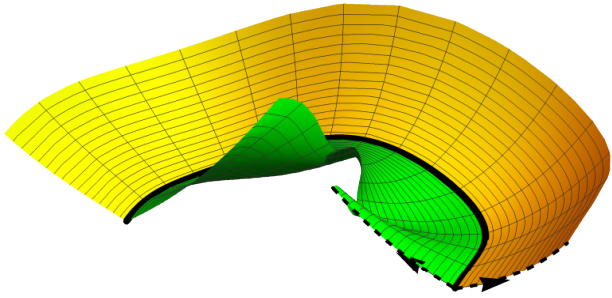

FIG. 10: The nonlinear isometries corresponding to $\delta_\pm = 0.1\ \cos(4\ \theta)$ and $\gamma_\pm = 0.2 + 0.15\ \cos(4\ \theta)$ joined continuously at u = 0. The region $u < 0$ represents the surface deformed by the isometry corresponding to $Y_{z-}$ which satisfies $\kappa_N < 0$. The $u > 0$ region in green (shaded), corresponds to $Y_{z-}$ and satisfies $\kappa_N > 0$. The dashed curves satisfy $\theta = 0$, and the arrows are their tangents at u = 0. As explained in the text the arrows are perpendicular to the rigidifying curve and must stay strictly above the xy-plane becuase of the requirement $\gamma_\pm > 0$.

## IV. CONCLUSION

We showed in section II C that some of the infinitesimal isometries of surfaces diverge near a rigidifying curve. Taken at face value this seems to indicate a reduction in the number of isometries of the surface near these curves. Indeed it can be shown using geometric arguments (see appendix A and [1]) that rigidifying curves have constant curvature $\kappa$ under linear isometries.

On the other hand, the experiments in [12] show that folding along curves with $\kappa_N = 0$ can happen continuously without a stretching energy barrier. We have shown, in section III, how the presence of nonlinear terms in the isometry equations reconciles these two observations. The argument for the rigidity of the $\kappa_N = 0$ curves relies on the assumption that an expansion of the form $\mathbf{X}(\epsilon) = \mathbf{X}_0 + \epsilon\ \mathbf{X}_1 + \cdots$ exists, where $\epsilon$ parametrizes the isometries. However we have shown, using series and boundary layer approximations of the full isometry equations that the solution cannot be analytic in $\epsilon$. In the full nonlinear solution, the normal curvature can be different from zero after deformation.

Moreover, we found pairs of solutions having opposite signs of normal curvature across a rigidifying curve. These correspond to continuous solutions across $\kappa_N = 0$ curves.

Further work must be done to understand the energetics of these "folded" isometries and why they seem to be realized in experiment instead of the smooth isometries of Fig. 8. Since the smooth isometries have an unavoidably large component proportional to $u^2$ whereas the folded on can have arbitrarily small displacements, there will be a range in parameter space where the folded solutions- with a suitably smoothed fold- is favorable energetically. However it is likely that the smooth solutions are not realized because of a lack of a low energy paths in deformation space leading to them starting from the reference surface, even if their energy is lower. The folded solutions on the other hand can start infinitesimally close to the starting surface and be varied continuously.

We acknowledge valuable conversations with A. Evans, B. Chen and CS acknowledges the generous hospitality of the Kavli Institute of Theoretical Physics in Santa Barbara. This work was funded by the National Science Foundation under award EFRI ODISSEI-1240441.

## APPENDIX A: ISOMETRIES AND GEOMETRIC NONLINEARITIES

Here we will examine the nature of isometries from a general geometric perspective. This will provide us with guidance and a bird's eye view of what to expect for the isometry spectrum of a surface. The main two guiding principles will be Bonnet theorem [24] and the relationship between the normal and geodesic curvatures [20].

Let us start by considering the Gaussian curvature $K(g_{ij})$, which is a function of the metric. Gauss's theorema egregium states that

$$b_{12}^2 = b_{11}\kappa_N - K\rho^2, \tag{A1}$$

using Gaussian normal coordinates [25], we have $d\ell^2 = (du^1)^2 + \rho^2(u^1, u^2)(du^2)^2$ and where $\kappa_N = b_{22}/\rho^2$ is the normal curvature along lines of constant $u_1$.

Under an isometry, the last term, $K\rho^2$ must remain constant. Eq. (A1), together with

$$\partial_1 b_{12} = \partial_2 b_{11} - b_{12}\frac{\partial_1 \rho}{\rho} \quad \text{and} \tag{A2}$$

$$\rho\partial_1\left(\rho\kappa_N\right) = \partial_2 b_{12} + b_{11}\rho\partial_1\rho - b_{12}\frac{\partial_2\rho}{\rho},$$

form the Gauss-Codazzi-Mainardi (GCM) equations. Bonnet's theorem [24] states that if (A1) and (A2) are satisfied, a unique surface will be determined up to rotations and translations. Using this and the Cauchy Kowalevski (CK) theorem applied to the GCM equation, we can say something general about the local existence and number of isometries without restricting ourselves to infinitesimal isometries.

Consider the vicinity of an arbitrary curve on the surface. Without significant loss of generality we assume this curve satisfies $u^1 = 0$. The CK theorem states that, as long as all the coefficients on the right-hand side of (A2) are analytic, there will be a unique solution in the vicinity of the curve for an arbitrarily specified $b_{12}(u^1 = 0, u^2)$ and $\kappa_N(u^1 = 0, u^2) \neq 0$. In other words, the curve will have two isometric degrees of freedom as long as $\kappa_N \neq 0$ on the final deformed surface. However, $\kappa_N$ may well be vanishing on the starting surface, as in the case of a torus. This is consistent with the inner solutions we found in section III; we can specify $\delta_\pm$ and $\pm\sqrt{\gamma_\pm} \sim \kappa_N$ to determine the solution uniquely.

On the other hand, if at any point on the $u^1 = 0$ curve we have $\kappa_N = 0$ (on the final surface), the CK theorem fails and there is no guarantee of solutions. However, in this case, we can determine what happens by first expanding $b_{11} = b_{11}^{(0)}(u^2) + b_{11}^{(1)}(u^2)u^1$, $b_{12} = b_{12}^{(0)}(u^2) + b_{12}^{(1)}(u^2)u^1$, $\rho^2 K = K^{(0)}(u_2) + K^{(1)}(u^2)u^1$ and $\kappa_N = \kappa_N^{(1)}(u_2)u^1$, and collecting terms with common powers of $u^{(1)}$. We obtain

$$b_{11}^{(0)}\kappa_N^{(0)} - \left(b_{12}^{(0)}\right)^2 = K^{(0)} \quad (A3)$$
$$b_{11}^{(0)}\kappa_N^{(1)} + b_{11}^{(1)}\kappa_N^{(0)} - 2b_{12}^{(0)}b_{12}^{(1)} = K^{(1)}.$$

and so on. From Eqs. (A2), we see that

$$\begin{aligned} b_{12}^{(1)} &= \partial_2 b_{11}^{(0)} + b_{12}^{(0)}\kappa_g^{(0)} \\ \kappa_N^{(1)} &= \kappa_g^{(0)}\kappa_N^{(0)} + \partial_2 b_{12}^{(0)} - \kappa_g^{(0)}b_{11}^{(0)}. \end{aligned} \quad (A4)$$

where $\kappa_g = -\partial_1 \rho$ is the geodesic curvature of the $u^1 = 0$ curve and $\rho(0, u^2) = 1$. Putting this together, we obtain a constraint in terms of the intrinsic geodesic curvature and Gaussian curvature in the vicinity of the curve,

$$\begin{aligned} K^{(1)} &= b_{11}^{(1)}\kappa_N^{(0)} + 2b_{12}^{(0)}\left[\partial_2 b_{11}^{(0)} + b_{12}^{(0)}\kappa_g^{(0)}\right] \\ &\quad - b_{11}^{(0)}\left[\kappa_g^{(0)}\kappa_N^{(0)} + \partial_2 b_{12}^{(0)} - \kappa_g^{(0)}b_{11}^{(0)}\right]. \end{aligned} \quad (A5)$$

When $\kappa_N = 0$, Eqs. (A1) and (A5) turn into a constraint entirely on the boundary curve because $b_{11}^{(1)}$ drops out. This explains why the inner solutions are singular when $\sim \kappa_N^2 \sim \gamma_\pm = 0$: we can specify $b_{12}^{(0)}$ arbitrarily close to the point $\kappa_N = 0$, but not exactly on the point, this leads to a singularity in the solution which we see in the series solution to Eq. 27.

Note that the only isometry of a torus with $\kappa_N = 0$ everywhere on the rigidifying curve is the torus itself, this is easy to see because Eqs. (A1) and (A5) completely determine $b_{12}^2 = -\rho^2 K = 0$ and $b_{11} = -K^1/\kappa_g$, these in turn can be used to determine the full series solution in the variable $u^1$.

Now we turn our attention to linear deformations, general geometric arguments provide guidance here as well, and can shed light on what is special about surfaces with $\kappa_N = 0$ curves. Imagine a one parameter family of isometries $\mathbf{X}(\epsilon)$, where $\mathbf{X}(0)$ is the starting surface and $\mathbf{X}(\epsilon)$ is the final surface. A linearized isometry can be expressed as $d\mathbf{X}(\epsilon)/d\epsilon|_{\epsilon=0}$. For any curve on the surface $\mathbf{X}(\epsilon)$ we can write the following geometric identity

$$\kappa^2(\epsilon) = \kappa_N^2(\epsilon) + \kappa_g^2. \quad (A6)$$

The linearized version of this identity is $\kappa\,\dot\kappa = \kappa_N\,\dot\kappa_N$, where a dot over the symbol means a derivative with respect to $\epsilon$. On the rigidifying curve $\kappa_N = 0$, in this case it is obvious that for the linear isometry we have $\kappa = \kappa_g + O(\epsilon^2)$, implying rigid motion of the curve, without change in curvature. It can easily be checked, using Eq. (20), that the finite linear isometries to the parabolic torus do indeed satisfy this property.

Yet another check on our solution comes from Eq. A1. The linearized version of the equation is written as

$$0 = \dot b_{11}(0)\,\kappa_N(0) + \dot\kappa_N(0)\,b_{11}(0) - 2\dot b_{12}(0)\,b_{12}(0). \quad (A7)$$

On the rigidifying curve of the parabolic torus this gives $\dot\kappa_N(0) = 0$, implying that the normal curvature is zero in the linearized isometric deformation. In addition the diverging linear solutions are inconsistent in the linear regime because they have non-zero normal curvature. Yet as we have already seen, $\kappa_N \neq 0$ on the final surface is perfectly well behaved as a nonlinear isometry. Therefore the divergence in the linear solutions is only a reflection of the fact that $\mathbf{X}(\epsilon)$ is not analytic near $\epsilon = 0$.

To conclude this section we demonstrate the non-analyticity of $\mathbf{X}(\epsilon)$ using a simple argument. Eq. (A6) can be rewritten as

$$\kappa_N(\epsilon) = \pm\sqrt{\kappa^2(\epsilon) - \kappa_g^2}. \quad (A8)$$

The first order derivative with respect to $\epsilon$ diverges at $\epsilon = 0$. Indeed, expanding to first order gives $\kappa_N \approx \pm\sqrt{2\epsilon\,\kappa_g\,\dot\kappa(0)}$, which is inconsistent with a first order expansion $\mathbf{X}_0 + \epsilon\,\mathbf{X}_1$ and $\kappa_{N0} + \epsilon\,\kappa_{N1}$, thus explaining the appearance of singular solution in the linear regime.

### APPENDIX B: SERIES AND NUMERICAL SOLUTIONS OF ISOMETRIES

The aim of this appendix is to verify our inner solutions against a series and numerical solution of the full isometry equations (24). Though we will still neglect the nonlinearities in $A_\theta$ and $A_s$, we can verify explicitly using the approximate inner solutions that these terms are indeed subdominant.

We first check that the series solution of Eq. (27) is consistent with the inner approximate solution. Note that Eq. (27) is derived from Eq. (24) by eliminating $A_s$ and $A_\theta$ and then taking derivatives of the third equation. Therefore any solution of Eq. (24) is a solution of Eq. (27), but the converse is not true. In order to make sure that the solutions we find are consistent with the isometry equations, we check that we can use Eq. (24) with $A_z(R) \equiv \epsilon\, \delta$ and $\partial_s A_z(R) \equiv \pm\sqrt{\epsilon\, \gamma}$ to determine the series coefficients of $A_s$ and $A_\theta$. The parameter $\epsilon$ is introduced here to control order $\epsilon$ terms of the series solution. Thus any solution of Eq. (27) is consistent with Eq. (24) only with a particular choice of the integration constant $\tau(\theta)$, which was given to leading order in $\epsilon$ back in Eq. (44).

In terms $\delta(\theta)$ and $\gamma(\theta)$, the series solution gives

$$\partial_u^2 Y_z(0,\theta) = \frac{\pm\epsilon\, \gamma'_\pm\, \left(\pm\gamma'_\pm - 4\,\sqrt{\epsilon\,\gamma_\pm}\, \delta'_\pm\right) + 4\,\gamma_\pm\, \left(\epsilon^2\, \delta'^2_\pm - 2\,\left(\pm\sqrt{\epsilon\,\gamma_\pm} + \epsilon\, \delta''_\pm\right)\right)}{4\,\gamma_\pm\, \left(\pm\sqrt{\epsilon\,\gamma_\pm} + \epsilon\, \delta''_\pm\right)}, \tag{B1}$$

However, from the inner solutions we get $\partial_u^2 Y_z(0,\theta) = -2$. To leading order in epsilon, the two expressions agree. This happens at every order in u. We can use boundary layer theory (see [23]) to determine to which order (in $\epsilon$) the inner solution is valid for every term in the series (in u). This can be done by defining

$$Y_z(u,\theta) \equiv \epsilon\, \Upsilon(\mu,\theta) \quad \text{and} \quad \mu \equiv \frac{u}{\epsilon^{1/2}}, \tag{B2}$$

where $\Upsilon$ as well as it's derivatives are O(1) in the interior layer which has width of order $\epsilon^{1/2}$. Using this we can expand $Y_z(u,\theta)$ to get

$$Y_z(u,\theta) \approx \epsilon\, \Upsilon + \epsilon^{1/2} \partial_u \Upsilon\, u + \frac{\partial_u^2 \Upsilon\, u^2}{2} + \frac{\partial_u^3 \Upsilon\, u^3}{6\, \epsilon^{1/2}}. \tag{B3}$$

Hence we see that the term proportional to $u^2$ is O(1) with approximation error scaling as O($\epsilon^{1/2}$). This is indeed what we find in equation (B1). Using Mathematica we extend Eq. (B1) to find terms up to order $u^{11}$ in the expansion of $Y_z(u,\theta)$. Fig. (B1) compares the numerical and series solutions of (27) to the approximate solution that we obtain by combining the inner and outer(linear) solutions to form a global approximation.

Finally we use $\gamma_\pm$ and $\delta_\pm$ to find the series expansion of $A_\theta$ and $A_s$, up to integration in the $\theta$ direction. We confirm that we can indeed use $\gamma_\pm$ and $\delta_\pm$ to determine the isometry, which implies that solutions of (27) that we find are indeed consistent with solutions of (24). For example, from the series solution, we have that $\partial_u A_s(0,\theta) = -\epsilon\, R^2\, \gamma_\pm/(8\, a^2)$ which matches what we find from the inner approximate solution. We will not explicitly show the rest of the series solution here for brevity.

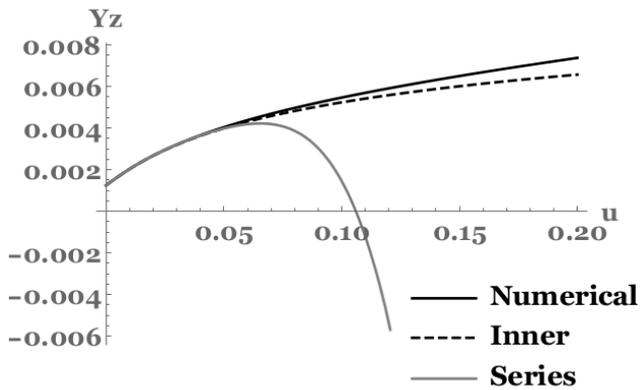

FIG. B1: Comparing the series (up to order $u^8$) and numerical solutions of Eq. (27) to the inner approximation. Here, a = 1, R = 3, $\delta_+ = 0.002 \cos(2\theta)$ and $\gamma_\pm = 0.006 + 0.003 \cos(2\theta)$.